\documentclass[]{raa}           
\usepackage{graphicx,times}
\usepackage{natbib}
\usepackage{amssymb,amsmath}
\bibpunct{(}{)}{;}{a}{}{,}

\usepackage{threeparttable}

\usepackage[a4paper=true,pagebackref=true]{hyperref}
\hypersetup{colorlinks = true, linkcolor = green, anchorcolor = red, citecolor = blue, filecolor = red, pagecolor = red, urlcolor = red}

\begin{document}

\title{Automatic removal of false image stars in disk-resolved images of the Cassini Imaging Science Subsystem}

\volnopage{ {\bf 20XX} Vol.\ {\bf X} No. {\bf XX}, 000--000}
\setcounter{page}{1}

\author{Qingfeng Zhang\inst{1,2}, Zhicong Lu\inst{1}, Xiaomei Zhou\inst{1}, Yang Zheng\inst{1}, 
  Zhan Li\inst{1,2}, Qingyu Peng\inst{1,2}, Shun Long\inst{1}, Weiheng Zhu\inst{1}}

\institute{Department of Computer Science, Jinan University, Guangzhou 510632, China; {\it tqfz@jnu.edu.cn}\\
        \and
        Sino-French Joint Laboratory for Astrometry, Dynamics and Space Science, Jinan University, Guangzhou, 510632, China\\
  \vs \no
   {\small Received 20XX Month Day; accepted 20XX Month Day}
}

\abstract{Taking a large amount of images, the Cassini Imaging Science Subsystem (ISS) has been routinely used in astrometry. In ISS images, disk-resolved objects often lead to false detection of stars that disturb the camera pointing correction. The aim of this study was to develop an automated processing method to remove the false image stars in disk-resolved objects in ISS images. The method included the following steps: extracting edges, segmenting boundary arcs, fitting circles and excluding false image stars. The proposed method was tested using 200 ISS images. Preliminary experimental results show that it can remove the false image stars in more than 95\% of ISS images with disk-resolved objects in a fully automatic manner, i.e. outperforming the traditional circle detection based on Circular Hough Transform (CHT)  by 17\%. In addition, its speed is more than twice as fast as that of the CHT method. It is also more robust (no manual parameter tuning is needed) when compared with CHT. The proposed method was also applied to a set of ISS images of Rhea to eliminate the mismatch in pointing correction in automatic procedure. Experiment results showed that the precision of final astrometry results can be improve by roughly 2 times than that of automatic procedure without the method. It proved that the proposed method is helpful in the astrometry of ISS images in fully automatic manner.
}

   \authorrunning{Q. Zhang et al. }            %author_head in even pages
   \titlerunning{Automatic Removal of False Image Stars}  % title_head in odd pages
   \maketitle

%________________________________________________ sections below
% 
\section{Introduction}           %% first-level sections will be auto-capitalized
\label{sect:intro}
Images from the Cassini Imaging Science Subsystem (ISS) have routinely been used to measure the astrometric positions of planetary satellites. \cite{Cooper06} reduced more than 200 ISS images of the Jupiter's satellites (Amalthea and Thebe). The satellites' center positions were computed by centroiding technique, and the pointing correction was performed automatically. The root mean square (RMS) residuals relative to the Jupiter's ephemeris JUP230 of the Jet Propulsion Laboratory (JPL) were 0.306 and 0.604 pixel for Amalthea and Thebe, respectively. \cite{Cooper08} described the discovery of Anthe and computed its orbit using the astrometry of ISS images. The Anthe's positions were measured using centroiding technique. The overall RMS fit residual was 0.456 pixel. \cite{Cooper14} reduced some ISS images of saturnian satellites' mutual events. Results showed that the measurement precision of inter-satellite separations between pairs of satellites within a given image was superior to that of the individual satellite. \cite{Tajeddine13, Tajeddine15} used the limb-fitting technique to measure the positions of several main starunian satellites from ISS images. The standard deviations of residuals were a few kilometers. \cite{Zhang17,Zhang18} have conducted astrometry of Enceladus using automatic method based on k-dimensional tree and Gaia DR1 star catalogue. Estimated standard deviations of residuals relative to JPL's SAT375 ephemeris were 2-4 kilometers. \cite{Cooper18} released one professional software named Caviar for reducing ISS images. These research activities indicate that the ISS images are an important resource for planetary astrometry.

\begin{figure}
\centering
\includegraphics[width=0.7\textwidth]{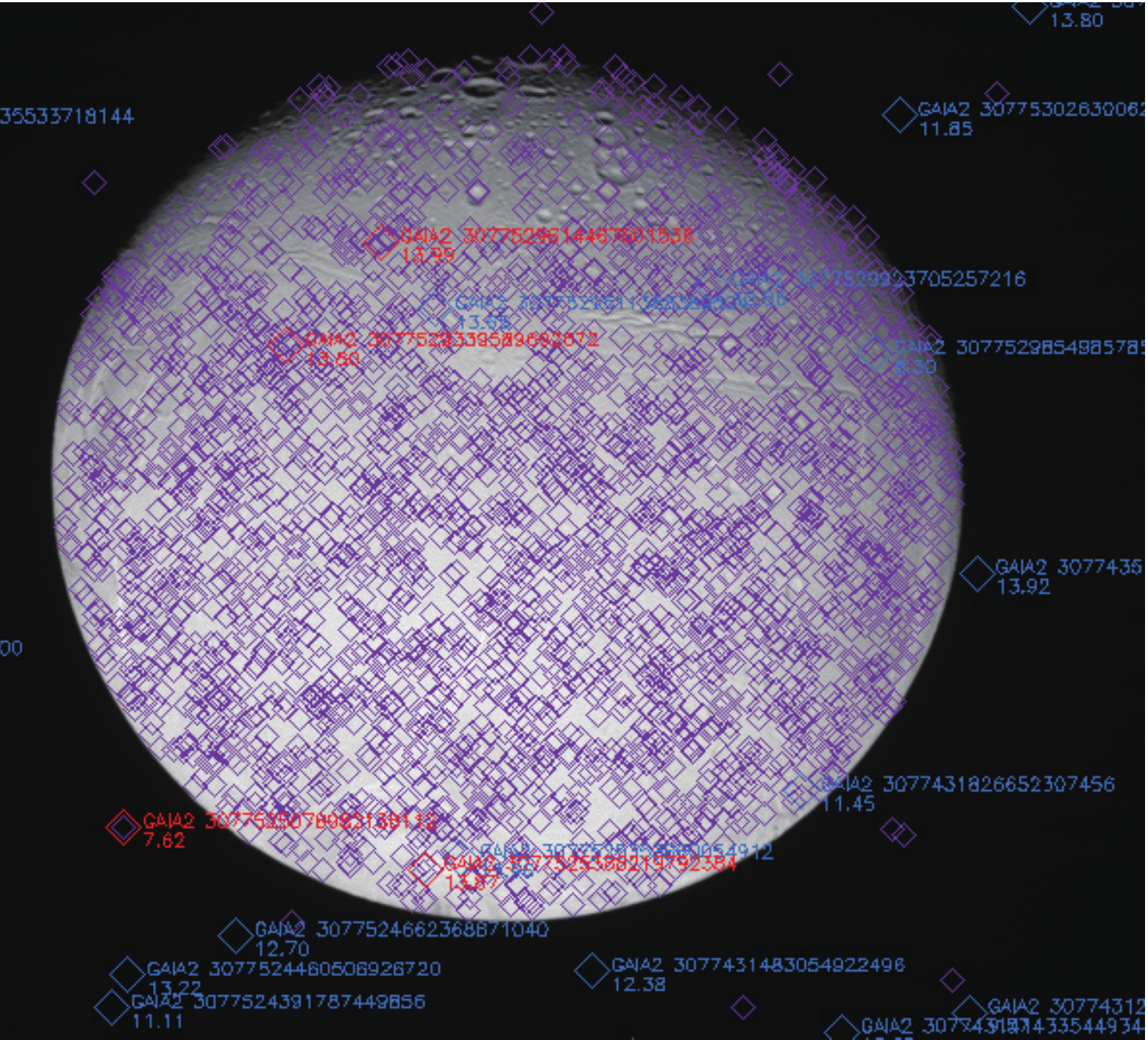}
\caption{The mismatching catalogue stars with image stars.}
\label{fig1}
\end{figure}

Since satellite targets are often observed closely by ISS, disk-resolved objects may appear in ISS images. Especially, during observations of mutual events, two disk-resolved satellites are often appeared in one image. It is necessary to match the detected image stars with catalogue stars to correct the camera pointing in the reduction of ISS images \citep{Cooper18}. However, false image stars are often detected in the disks, which may lead to the mismatching between image stars and catalogue stars, and the error of pointing correcting. As shown in Figure \ref{fig1}, there are three types of marks in the original ISS images: purple boxes, blue boxes and red boxes, which represent the detected image stars, catalogue stars in the same field, and the matched catalogue stars, respectively. The catalogue stars are also labeled by the text. The figure shows a lot of false image stars in the disk-resolved object. When the catalogue stars automatically match with the detected image stars, four matched stars are found (represented by the four red boxes). Obviously, three of them are false matched stars. These false matched stars should lead to errors of camera pointing correction. At present, the three mismatching stars can only be removed with a manual method. To enhance the efficiency and decrease the disturbance, automatic methods should be developed \citep{Zhang18}.

The key to remove false image stars is to extract the contour of disk-resolved objects. The contour extracting is often used in the limb-fitting technique in astrometry. \cite{Thomas95} measured the shape parameters of Vesta from images of Hubble Space Telescope (HST) using the limb-fitting technique. The contour is obtained by performing edge detection. The processing needs manual intervention. \cite{Peng03} measured the Jupiter images with the limb-fitting. They removed the halo of the Jupiter in an image, performed a four-neighbor detection to find the edge of the Jupiter, and used the circle fitting to determine the center position of the Jupiter. \cite{Mallama04} carefully studied the edge detection of the Jupiter in the HST images. Base on the assumption that the edge pixels were known, they used the interpolation method to determine the sub-pixel positions of the edge of the Jupiter and obtained the Jupiter's contour. \cite{Pasewaldt12} applied the limb-fitting to the Deimos's images taken by Mars Express. The Deimos's shape was fitted with the edge pixels to obtain its center position. The above methods can not be performed automatically.

\cite{Zharkova03} used the Canny operator to detect edges of the Sun and then extract its shape. \cite{Yuan17} applied methods, such as median filtering, binarization, and Hough transform, to determine edges of the Sun. These methods are suitable to extract the full-disk contour of the Sun. Nevertheless, in ISS images, the satellites most frequently appear partial disks and even crescent shapes. These automated methods can not be applied to ISS images.

In addition, all the above methods do not consider special cases of two disk-resolved objects in one image. Therefore, the objective of this study was to develop an automatic method to remove the false image stars in the disk-resolved objects in astrometry of ISS images.

\section{Method}
\label{sect:method}

\begin{figure}
\centering
\includegraphics[width=0.60\textwidth]{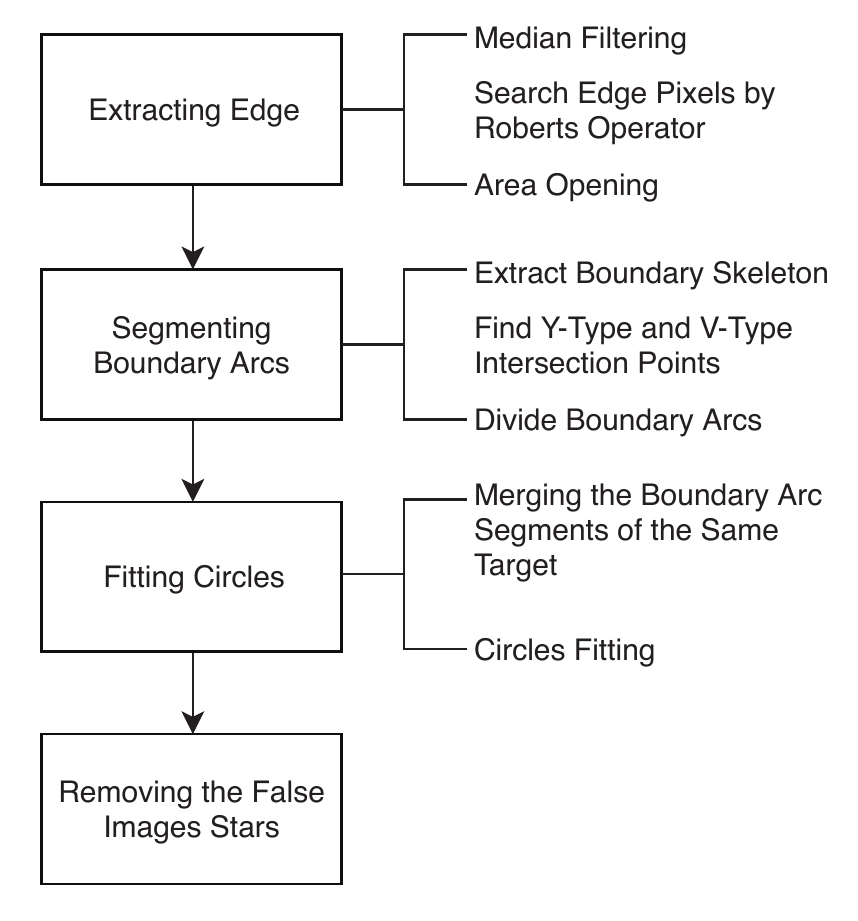}
\caption{The processing flow of our method, including four main steps.} \label{fig2}
\end{figure}

As shown in Figure \ref{fig2}, the propsed method includes the following steps: extracting edge, segmenting boundary arcs, fitting circles, and removing false image stars. The aim of extracting edge was to get boundary arcs of the disk-resolved objects. The boundary arcs segmentation split the special boundary arcs into two or three dependent arc segments belonging to only one disk-resolved object. In the fitting circles step, the boundary arc segments were fit into one or two circles. Finally, the false image stars were removed in the fitting circles. The detail procedures were described as follows.

\subsection{Extracting Edge}

\begin{figure}
\centering
\includegraphics[width=0.88\textwidth]{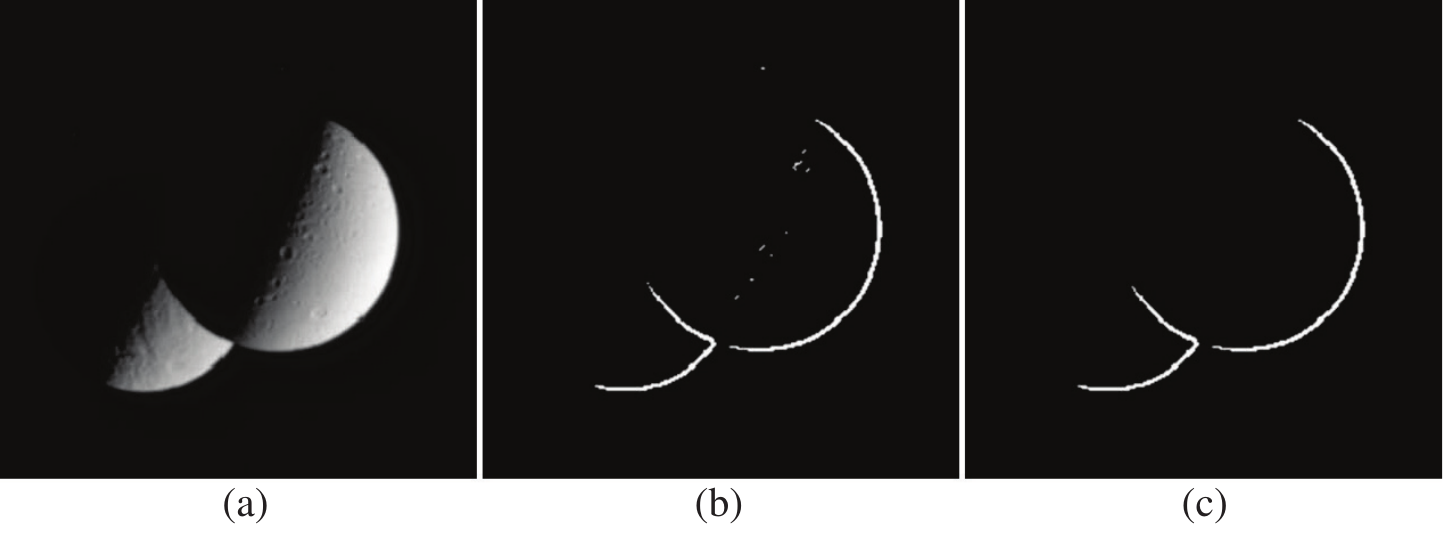}
\caption{A set of graphics through extracting edge.} \label{fig3}
\end{figure}

First, a median filtering of mask size of $2\times2$ was applied to reduce noise in ISS images as much as possible. Second, Robert operator was used to detect edge pixels and the images were converted  into binary ones with a threshold of 30\% of maximum gray. Figure \ref{fig3}(b) shows an example of the result after the two steps from the original ISS images (Figure \ref{fig3}(a)). Obviously,  some unwanted edge pixels still exist in Figure \ref{fig3}(b).

Finally, the area opening was used to reduce the redundant edge pixels, which revealed the real contour pixels according to the connectivity \citep{Vincent93}. The edge pixels constructure some connected sets (CS hereafter). In this step, the edge pixels in the CSs with number of pixels $\geqslant$ a threshold $T$ (to be determined as follows) were kept and the other CSs were removed.

Given $N_i$ is the number of pixels in $i^{th}$ CS, all $N_i$ are sorted in descending order. The threshold $T$ can be determined by the first $N_i$ which satisfy the equation below:
\begin{equation*}
T= N_i,\quad if\quad \frac{N_i}{N_{i-1}}>6.0\quad (i=n, n-1, \cdots, 2)
\end{equation*}

Where 6.0 is an empirical number, which is determined by experiments.

After the area opening, edge pixels in the larger CSs were preserved and isolated small groups of edge pixels were removed (Figure \ref{fig3}(c)).

\subsection{Segmenting Boundary Arcs}

\begin{figure}
\centering
\includegraphics[width=0.88\textwidth]{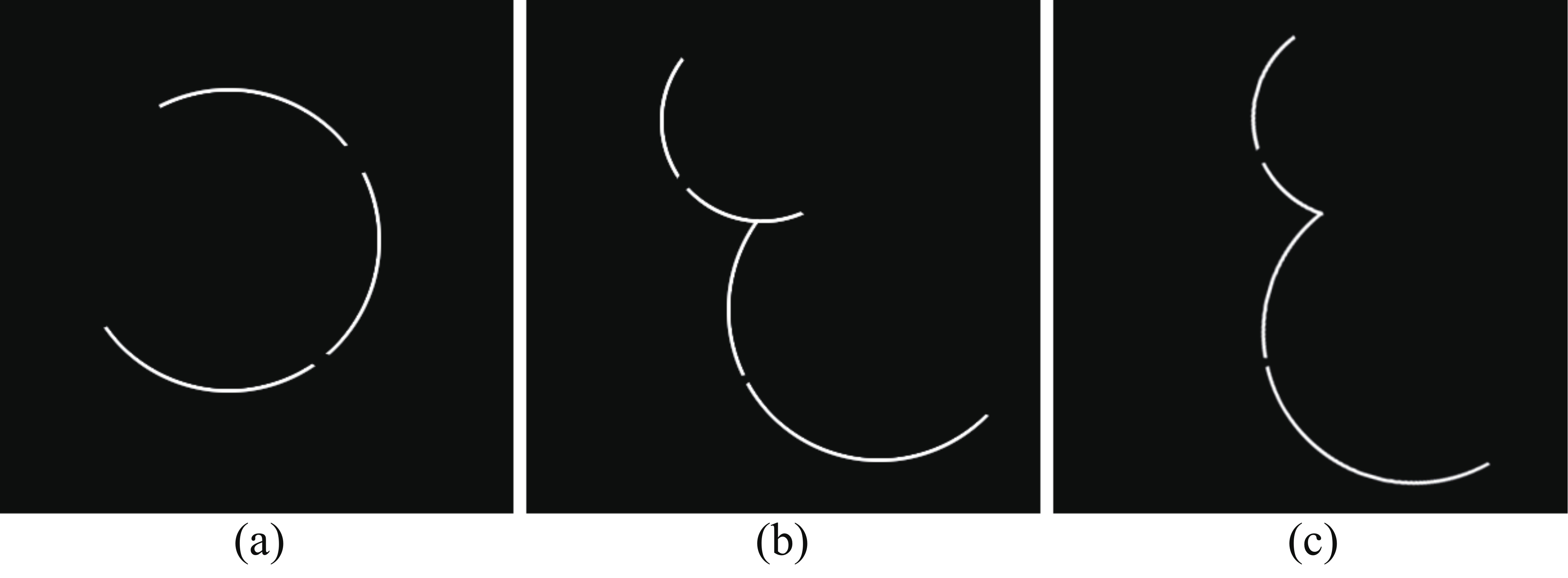}
\caption{Boundary arc segment and intersection points. (a) One boundary arc contains several arc segments. (b) The boundary arcs contain one Y-type intersection point. (c) The boundary arcs contain one V-type intersection point.} \label{fig4}
\end{figure}

After extracting edges, one or two boundary arcs may be composed of several CSs. Figure \ref{fig4}(a) shows that one arc is composed of several CSs. Figure \ref{fig4}(b) shows the case containing a Y-type intersection point. In the case, several CSs form two arcs. One of them includes a Y-type intersection point of two arcs. Figure \ref{fig4}(c) is similar to Figure \ref{fig4}(a), but one V-type intersection point is included in the CS involving two objects. The CSs with Y-type or V-type intersection points need to be divided to make each CS belong to only single object. 

Boundary arc segmentation was used to divide the CS with Y-type or V-type intersection points.

Before dividing CS, skeleton extraction is necessary, which is performed by the iterative algorithm \citep{Haralick92}. The algorithm extract the skeleton of the CS and make the thick of each skeleton  with one pixel.

After extracting skeleton, each skeleton (one boundary arc segment) contains its endpoints and possible Y-type or V-type intersection points. If containing an intersection point, the CS is divided at the point. Otherwise, it is kept. Eventually, each CS belongs to a single object. The key to divide CS is to find the Y-type or V-type intersection points in the CS’s skeleton.

\begin{figure}
\centering
\includegraphics[width=0.36\textwidth]{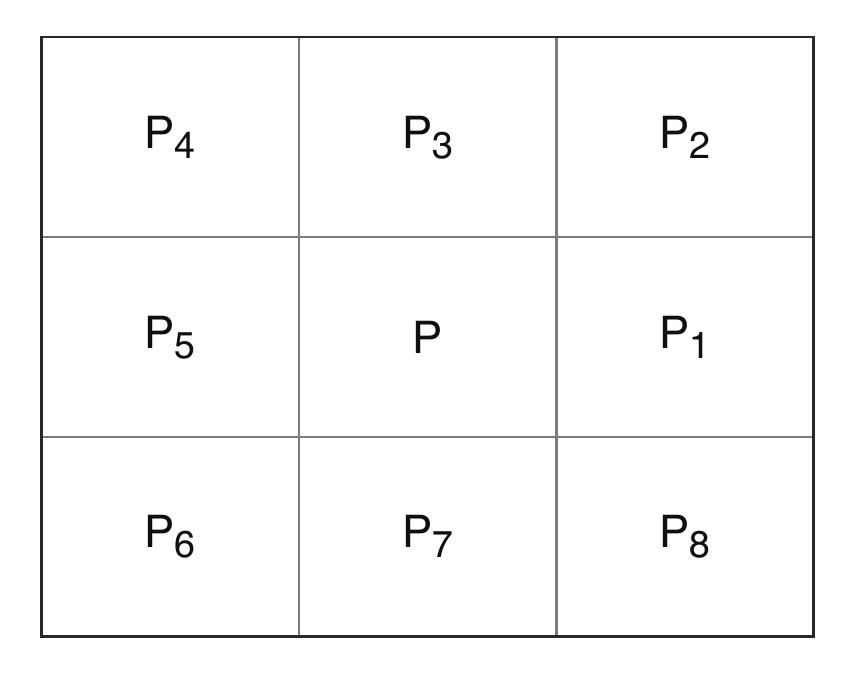}
\caption{A pixel P and its 8 neighboring pixels.} \label{fig5}
\end{figure}

First, the crossing number of a pixel was used to find the endpoints and Y-type intersection point in one CS's skeleton. This is a classical method applied in fingerprint identification \citep{Mehtre93}. Each pixel $P$ has 8 neighboring pixels in its $3\times3$ neighborhood (Figure \ref{fig5}). In the binary image of one-pixel thick skeleton, each pixel takes a value of either 0 or 1. The crossing number of pixel $P$ is computed as follows:
\begin{equation*}
c = \sum_{i = 1}^8 \lvert P_i - P_{i+1}\rvert \quad where \quad P_9=P_1
\end{equation*}

The pixel $P$ is endpoint with $c=2$ and Y-type intersection with $c=6$. 
The skeleton of a connected set was scanned to search its endpoints and Y-type intersection point, which were labeled.

Second, the Chord-to-Point Distance Accumulation (CPDA) technique was performed to search and label the V-type intersection points \citep{Awrangjeb09}. Finally, the CS was divided at the Y-type or V-type intersection points. With a Y-type point, the CS was divided into three sets. With a V-type point, the CS was divided into two sets. Actually, a CS has only one either Y intersection point or V intersection point. The division was implemented by zeroing the $5\times5$ neighborhood of the Y or V points. The operation was applied to every set to ensure that every CS belonged to only one object. The two original endpoints in every CS's skeleton were marked as 1-type endpoints. The endpoints generated from Y and V-type points were marked as 2-type endpoints.

\subsection{Fitting Circles to the Segmented Arcs}

Since there are probably several CSs (i.e., arc segments) in one object's boundary, these arc segments belonging to the same object should be merged before fitting circles. The merge arc segment procedure included the following steps:

\begin{enumerate}
\item Pick out the longest arc segment S in the entire boundary.
\item Find the nearest arc segment C of S, merge it into S, and update the endpoint of S with the far endpoint in C. The nearest arc segment refers to the one that has the smallest distance and less than 7 pixels. The distance was counted from the 1-type endpoints of S and C. Eventually, the arc segment S was extended. Although a little gap existed between the endpoints of the original set S and C before extending, it was assumed that the extended set S was a big connected set after extending.
\item Repeat step 2 till there is no the nearest arc segment or the nearest arc segment's far endpoint is 2-type endpoint. 
\item Repeat step 1 to step 3 till there is no arc segment needing to be merged.
\end{enumerate}

In the end, the segmented arcs belonging to one single object were combined as much as possible. They form a set of new arc segments. The number of arc segments was decreased.

After merging segmented arcs, the least square circle fitting was applied to the longest large arc segment. The radius of the fitting circle was set as $R$. The average value of the five maximum distances between the circle center and the pixels in one certain arc segment was evaluated. If the averaged distance was $> 0.1R$, the certain arc segment doesn't belong to the fitting circle. Since there are at most two disk-resolved objects in one image, the arc segment should belong to the other object. With the method, the arc segments belonging to the other circle were picked out. Up to our knowledge, there is no ISS image which contains more than two disk-resolved objects. This is probably due to the narrow field of view of the narrow angle camera ($21'\times21'$). In addition, the image with more than two disk-resolved objects, if not absent, should be very rare. Base on the observations, processing this kind of images is not considered in this paper.

Finally, one or two circles were obtained to cover all arc segments. These circles presented the contour of the one or two disk resolved objects in the ISS images.

\subsection{Removing the False Image Stars}
It was assumed there were two fitting circles, i.e., $f_1(x,y)=0$ and $f_2(x,y)=0$. The image star with the coordinate $(x_s, y_s)$ satisfying the following condition is removed:
\begin{equation*}
f_1(x_s,y_s)\leqslant 0\quad or\quad  f_2(x_s,y_s)\leqslant 0
\end{equation*}

\section{Experiments and Comparison}
\label{sect:experiments}

\subsection{Experiments}

The automatic removal of false image stars (ARFIS hereafter) proposed in this paper was tested using 200 ISS images. Each image in the data set contains one or two disk-resolved objects, mostly two, which are more difficult to analyze. The image set covers different lighting conditions, such as very bright, medium bright, and dim disks. The image set also includes several positional relationships between two objects, such as in separated, close, and interleaved positions. In addition, various disk sizes are covered, such as the large, middle, and small disks. These 200 images basically represent all situations in the processing of Cassini images.

Figure \ref{fig6} shows some typical results. The original ISS images and the detected image stars marked by blue boxes in the original images are shown in Figure \ref{fig6}(a) and (b), respectively. Figure \ref{fig6}(c) presents the contours of disk-resolved objects detected by ARFIS. Figure \ref{fig6}(d) shows the fitting circles (in blue) superimposed on the original images. Figure \ref{fig6}(e) displays the final result with removing the false image stars in the disk-resolved objects.

In the 200 images, 191 images yielded correct circular contours, with an accuracy $> 95\%$. The errors were mainly from small and/or faint object's images. For the extremely faint objects, the ARFIS might not be able to catch the contours. Figure \ref{fig7}(a) shows that there are two objects and the fainter object's contour is missed. In practice, the object’s images are fainter, it is less possible to detect image stars in the object's limb. Hence, the missed contour of fainter object hardly influence the astrometry of the ISS images. For small objects, the ARFIS might detect contours that do not match with the real. As shown in Figure \ref{fig7}(b) there is a small disk-resolved object in the small red box. We magnify it to show its image and its detected contour in the left top sub image. From the magnified local image, it can be found that the detected contour is incorrect. But, the real area of the object's image is so small that it is little possible there are image stars in it. Hence, the incorrect contours of small objects also hardly influence the astrometry of ISS images. In sum, the errors of detected contour of faint or small objects have little effect on the astrometry of ISS images.

The ARFIS method was implemented in Matlab R2016a, running on a PC with Intel i5-3570 CPU, 16 GB memory and Windows 10 operation system. The average processing time of one image with the size of $1024\times1024$ is 0.173 seconds. The method is completely automatic.

\subsection{Comparison with Circular Hough Transform Method}

As described before, the circle detection is the key step to remove false detections of image stars. As an excellent method, circle detection based on Circular Hough Transform (CHT) has been widely used \citep{Duda72}. Thus, the circle detection with our method was compared with that based on CHT. Our method outperforms CHT in the following aspects: 

\begin{figure}
\centering
\includegraphics[width=0.98\textwidth]{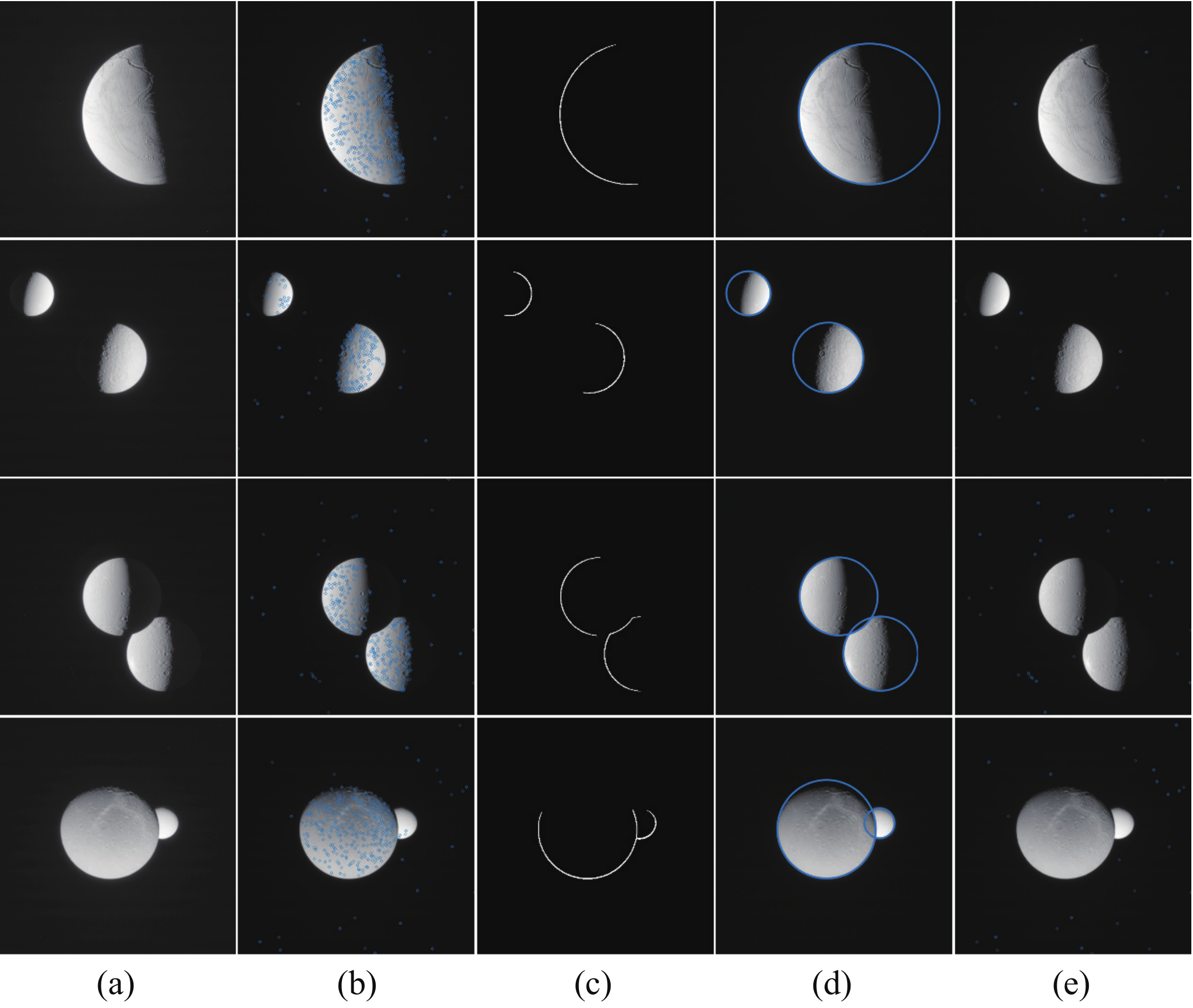}
\caption{Typical results from the proposed method: (a) Original images, (b) detected image stars in the original images, (c) detected contours, (d)fitting circles, and (e) final results.} \label{fig6}
\end{figure}

\begin{figure}
\centering
\includegraphics[width=0.78\textwidth]{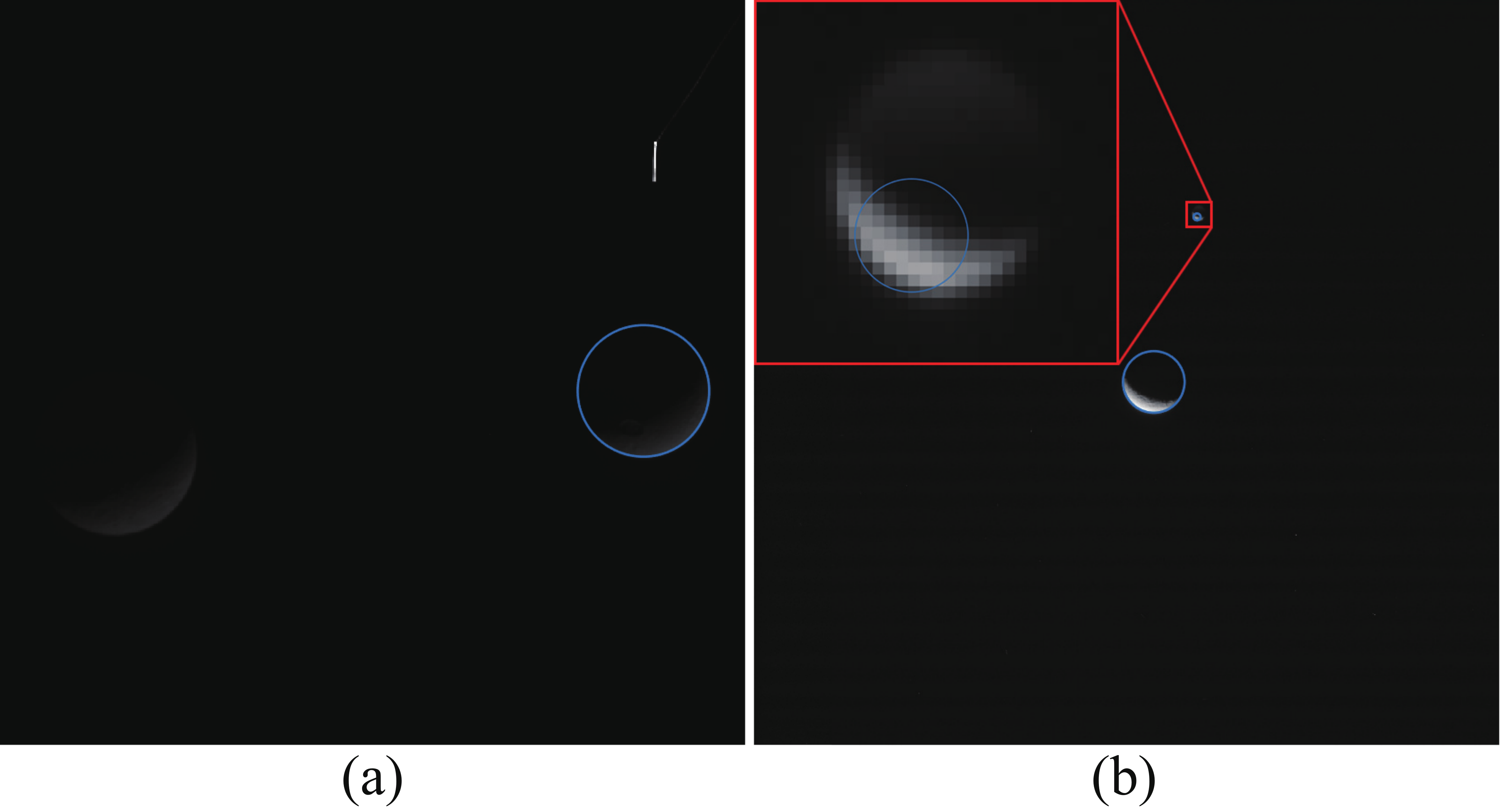}
\caption{The example of wrong contour detection. (a) Missing contour of the fainter object. There are two disk-resolved objects in the image, one object's contour is detected (marked by blue circle), the other is missed. (b) Incorrect contour of small object. The small object's contour is detected, but its contour is incorrect. The small object is magnified to show the error in the left top sub image.} \label{fig7}
\end{figure}

\begin{enumerate}
\item ARFIS can work for different images without manual parameter tuning. CHT is a method of indirectly finding circle, and its performance largely depends on the proper setting of parameters. However, every image has different number of objects and overlap situation between objects, and every object has different size, illumination conditions, and brightness. Therefore, it is difficult (if not impossible) to set the uniform parameter values to deal with the problems in most cases.

\item The accuracy of the fitting results with ARFIS is better than that with CHT. CHT does not find directly the center and radius of circle. The accuracy of finding circle depends on parameter setting, which was affected by noises and brightness distributions in images. The 200 ISS images were also tested with the CHT method. With carefully optimizing parameter setting, available results were obtained from 157 images, including approximated circular contours as shown in Figure \ref{fig8}(a-b). Figure \ref{fig8}(a-b) shows the circular contours (in blue) obtained by ARFIS are more correct than those (in red) from CHT method. The errors from CHT mainly presented in cases such as false circle detection, missing circle detection, and the detected contour mismatching with actual contour (Figure \ref{fig8}(c-e)).

\begin{figure}
\centering
\includegraphics[width=0.96\textwidth]{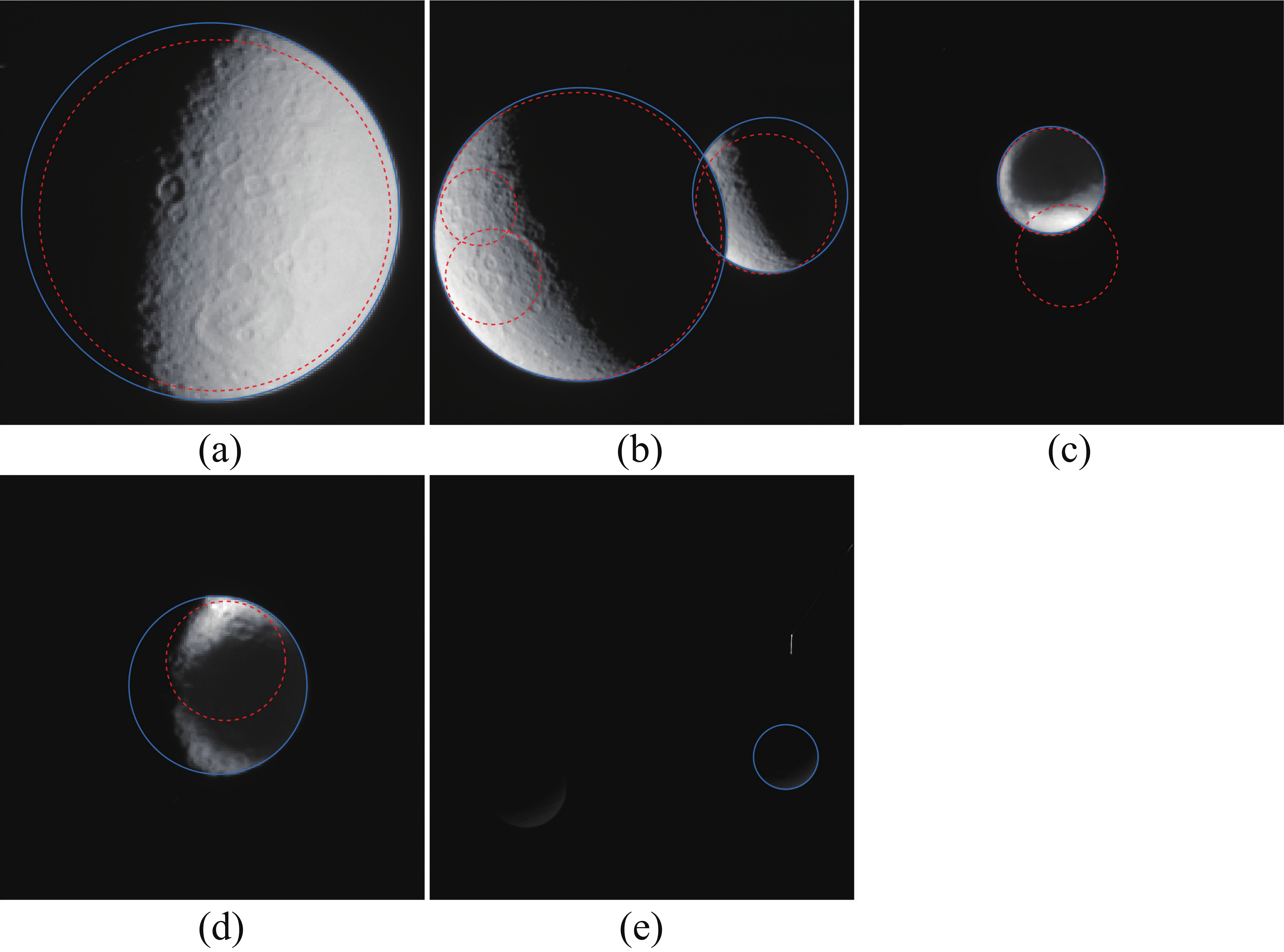}
\caption{Comparison between our method and the CHT method. (a) Available results were obtained from both methods, but our result (in blue) provided higher accuracy than CHT (in red). (b) Two accurate circular contour (in blue) were obtained from our method, while two approximate circular contours and two additional false circles (in red) were obtained from the CHT. (c) Correct detection (in blue) was obtained from our method, while an additional wrong detection (in red) was obtained from the CHT. (d) Our method provided correct detection, while CHT's detection did not match largly with the real. (e) Our method only missed one object's contour, while CHT did not detect any contour.} \label{fig8}
\end{figure}

\begin{figure}
\centering
\includegraphics[width=0.7\textwidth]{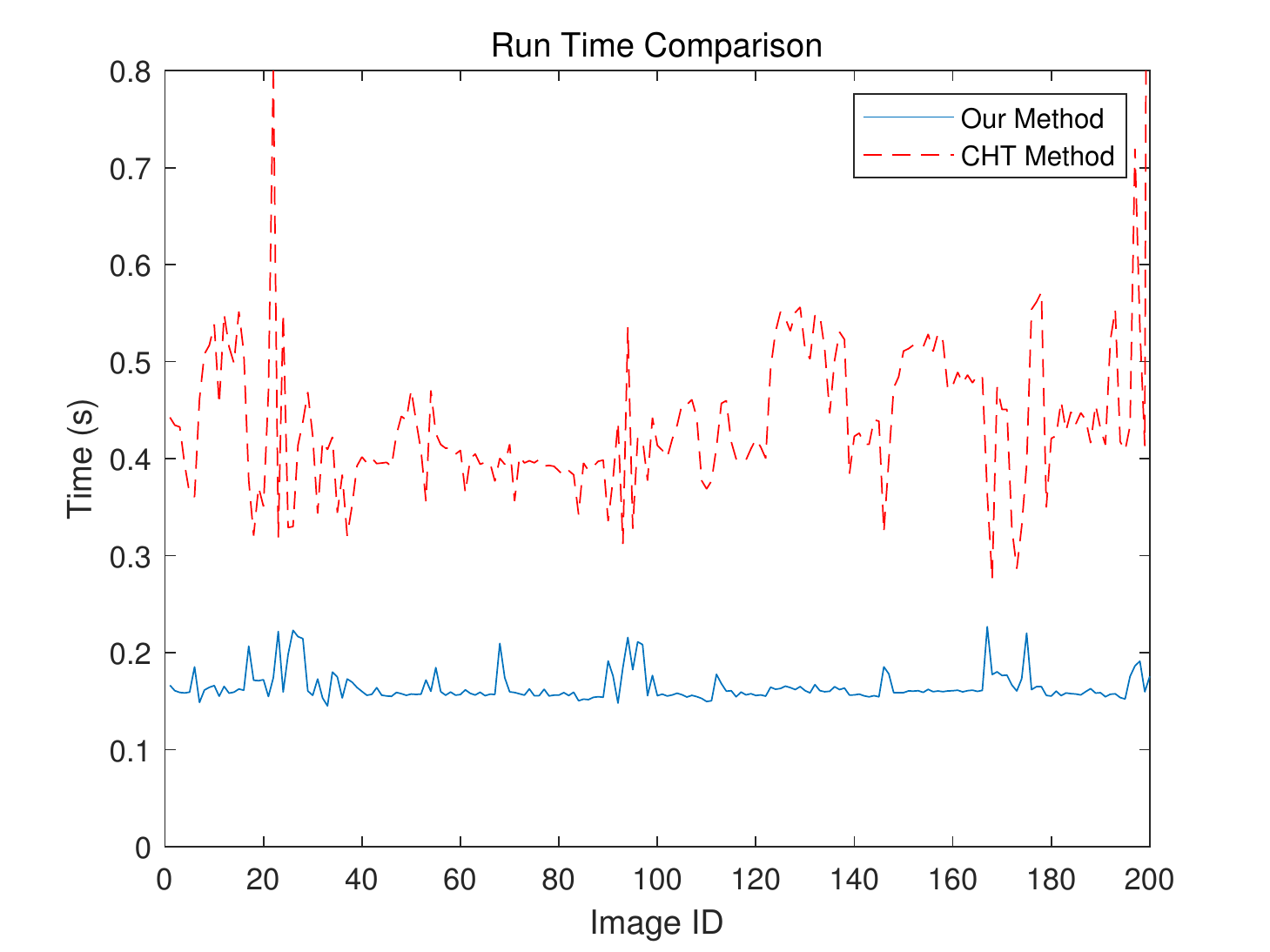}
\caption{Run time comparison.} \label{fig9}
\end{figure}

\item The calculation efficiency of ARFIS is higher than that of CHT. The highly optimized CHT based algorithm in Matlab toolbox was used to test the 200 ISS images. Every image's run time is displayed in Figure \ref{fig9}. Obviously, the running speed of ARFIS is more than twice as fast as that of the CHT method. It is because the CHT is computationally expensive. In ARFIS, the key step is using least square fitting to compute the center position $(a, b)$ and radius $r$ of the fitting circle. Its time complexity is $O(s)$, where $s$ is the number of contour pixels in an image. But, its counterpart in CHT is a voting computation in 3D space of $(a, b, r)$ whose time complexity is $O(m\times n\times s)$, where $m$ and $n$ are the size of an image. From this view, our method is faster than CHT. The experiment result is consistent with the analysis.
\end{enumerate}

\section{Applications}
As explained above, the false image stars introduced by the disk-resolved satellites in Cassini ISS images have a negative impact on accurate pointing correction. ARFIS can be used to eliminate false image stars and therefore significantly improve the accuracy in automatic method of astrometry of ISS images.

A quantitative comparisons has been made on a set of 32 ISS images of Rhea that is the target of measurement. Each image contains the disk-resolved Rhea. Since they are all images of Rhea and share the same ephemeris. Therefore there is no the influence of different ephemeris in the final observed-minus-computed (O-C) residuals. Experimental results suggest that mismatches may occur if ARFIS was not applied. However, once it was applied, all false image stars in disk-resolved objects were eliminated and an exact match was achieved.

\begin{figure}
\centering
\includegraphics[width=0.7\textwidth]{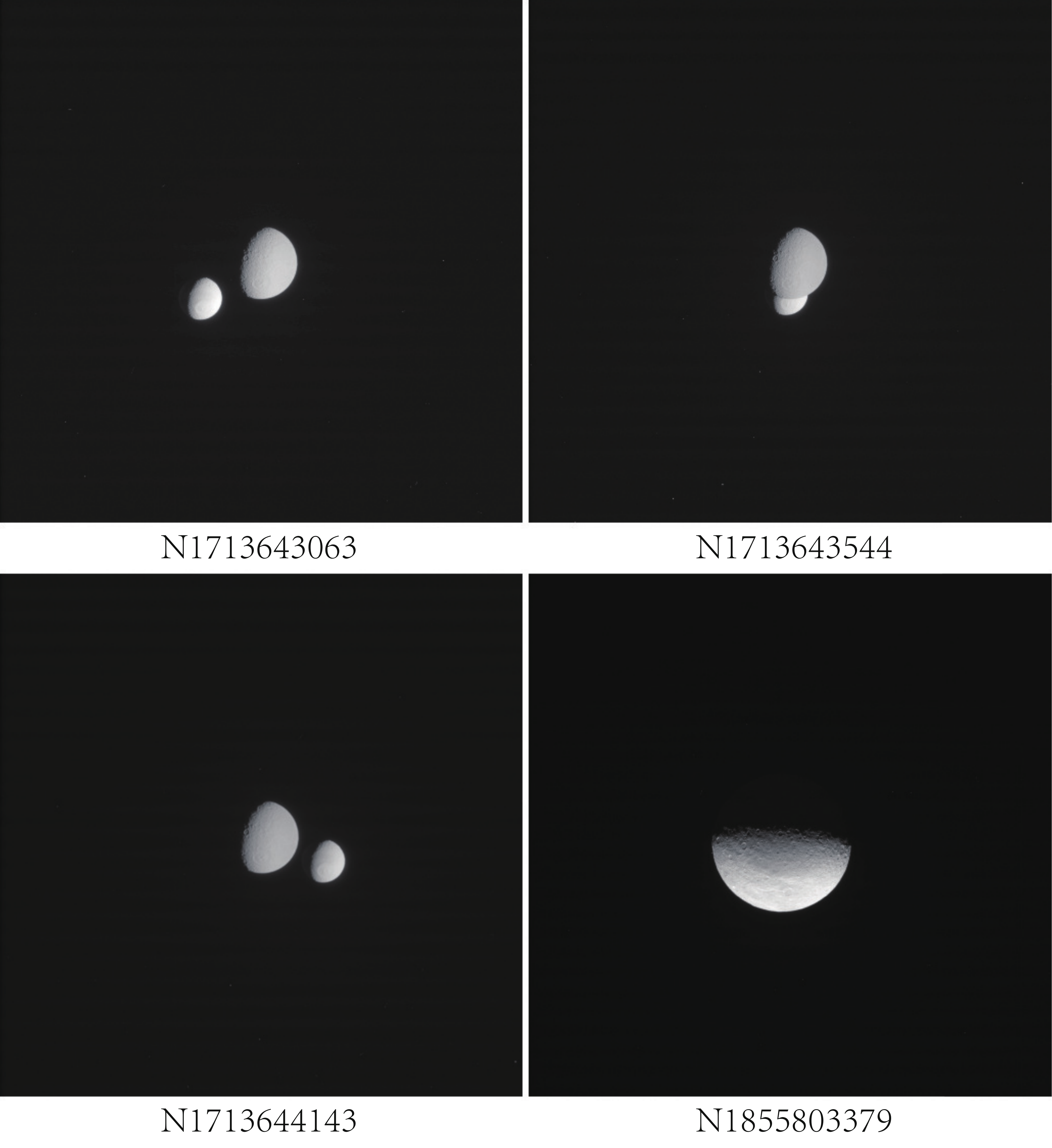}
\caption{Some typical ISS images of Rhea. The big disk-like object in these images is Rhea.} \label{fig10}
\end{figure}

Figure \ref{fig10} demonstrates typical images in the Rhea image set, in which either one or two disk-resolved satellites had been observed (in the latter cases, the two satellites are either far apart or interleaving). These satellites introduced false image stars, which in turn introduced mismatches with catalogue stars.

\begin{figure}
\centering
\includegraphics[width=0.9\textwidth]{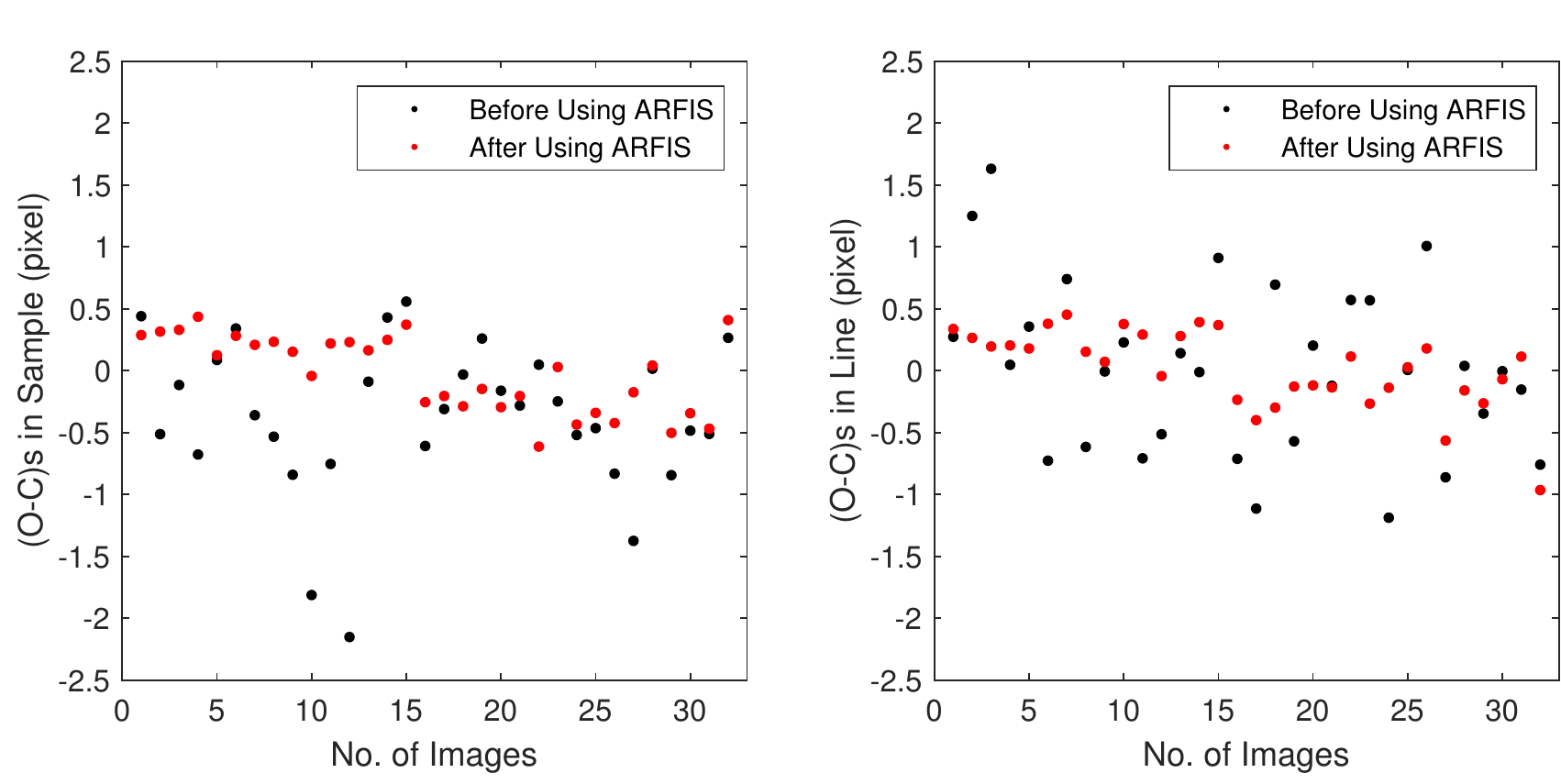}
\caption{Comparisons between before and after using ARFIS for these 32 images. (a) The pixel-wise (O-C)s in sample direction. (b) The pixel-wise (O-C)s in line direction.} \label{fig11}
\end{figure}

In the experiments, Caviar was used to measure the position of Rhea, first without ARFIS, and then the measurement was repeated, but with ARFIS. Comparisons were then made between these two sets of results and the results are shown in Figure \ref{fig11}, where (a) and (b) show the pixel-wise (O-C)s in sample and line, respectively. Here, we use the Cassini ISS convention of referring to the image pixel coordinate along the x axis as “sample” and the y coordinate as “line”. It should note that the images set used in experiment contains two parts. One part is the ISS images of mutual events of Rhea and Tethy taken on 2012-111T. The other part is the ISS images of Rhea taken on 2016-298T. The total number of images is 49. But 10 of these images are not available for astrometry because of no reference star in images. Furthermore, 7 of 39 images have the same results when measure with/without ARIFS, because they have the same reference stars no matter if ARFIS was applied. The remaining 32 images show the difference between using or not using ARFIS. Figure \ref{fig11} displays the results from the 32 images.

Table \ref{tab1} summarizes the O-C results of these 32 images, which shows that the results of reduction with ARFIS are better than those of without ARFIS. The residuals of observed position of Rhea relative to the JPL SAT389 were computed. The means of residuals after using ARFIS are close to 0 while there exists a non-negligible offset along sample direction for that before using ARFIS. The standard deviations in sample and line are 0.311 and 0.317 pixel respectively in the cases with ARFIS-enhancement, and 0.613 and 0.689 pixel in the cases without ARFIS-enhancement. This suggests that exact match yields roughly 1/2 of the standard deviation errors of mismatch.

\begin{table}
\centering
\caption{Comparison between the results from before using ARFIS and that from after using ARFIS. Mean values (Mean) and standard deviations (SD) of residuals of all of 32 observed positions of Rhea relative to the JPL SAT389 ephemeris in sample and line (in pixels).}
\label{tab1}
\begin{threeparttable}
\begin{tabular}{ccccc}
\hline
&\multicolumn{2}{c}{Sample\tnote{*}}&\multicolumn{2}{c}{Line\tnote{*}}\\
&Mean&SD&Mean&SD\\
\hline
Before Using ARFIS&-0.378&0.613&0.007&0.689\\
After Using ARFIS&-0.021&0.311&0.018&0.317\\
\hline
\end{tabular}
\begin{tablenotes}
\footnotesize
\item[*] According to the convention of ISS images, the image coordinate system is described by sample and line. Its origin is at the center of the top left pixel, with line increasing downwards and sample to the right, when the image is displayed in its normal orientation.
\end{tablenotes}
\end{threeparttable}
\end{table}

The above results suggest that the ARFIS we propose in this paper can significantly improve the accuracy and precision of astrometry. It is worth noting that this improvement is achieved in automatic manner. Exact match will occur when it comes to manual processing, because the false match between image stars and catalogue stars can be manually removed. However, ARFIS is still applicable in the manual case, in that it helps to save the efforts of manual mismatches elimination, efficiency can therefore be improved. 

\section{Conclusion}
\label{sect:conclusion}

In the paper, a method was developed to remove false image stars in disk-resolved objects for ISS images. The method was automatic. Compared to the classical Circular Hough Transform, the method can remove the false image stars in more than 95\% of ISS images with disk-resolved objects in a fully automatic manner, i.e. outperforming the CHT by 17\%. In addition, its speed is more than twice as fast as that of the CHT method, and it can handle more cases without manual parameter tuning. When the method is applied in astrometry of ISS images, the precision of final astrometry results can be improve by roughly 2 times than that of automatic procedure without it. Experiments proved that the proposed method is helpful in the astrometry of ISS images in fully automatic manner.

\clearpage

\normalem
\begin{acknowledgements}
We thank the anonymous reviewer whose comments and suggestions helped substantially improve this manuscript. This work was partly supported by National Natural Science Foundation of China (Grant No. U1431227, 11873026), Natural Science Foundation of Guangdong Province, China (Grant No. 2016A030313092), National Key Research and Development Project of China (Grant No. 2019YFC0120102), and the Fundamental Research Funds for the Central Universities (Grant No. 21619413).
\end{acknowledgements}

\bibliographystyle{raa}
\bibliography{biblio-ARFIS}

\end{document}